\documentclass{PoS}
\usepackage{epsfig}
\usepackage{amsmath}
\usepackage{braket}

\title{Stochastic All-to-All Propagators for Baryon Correlators }

\ShortTitle{Stochastic All-to-All Propagators for Baryon Correlators }

\author{\speaker{John Bulava}\hspace{1mm}$^a$\thanks{Email: jbulava@andrew.cmu.edu}\hspace{1mm}, Robert Edwards$^b$\thanks{Email: edwards@jlab.org}\hspace{1mm} and 
Colin Morningstar$^a$\thanks{Email: colin\_morningstar@cmu.edu}\hspace{1mm}, for the Hadron Spectrum Collaboration \\
\llap{$^a$} Department of Physics, Carnegie Mellon University, Pittsburgh, PA 15213, USA  \\
\llap{$^b$} Thomas Jefferson National Accelerator 
Facility, Newport News, VA 23606, USA \\}

\abstract{The effectiveness of various dilution schemes in the evaluation of baryonic two-point functions is compared. The error of a representative set of observables as a function of the number of Dirac matrix inversions is used as a basis for comparison. To achieve an equivalent reduction in error, we demonstrate that an increase in the number of dilution projectors on a single noise source usually requires fewer inversions than the use of multiple noise sources.  This exploratory study was performed on 100 quenched gauge configurations and will be applied to the calculation of low-lying hadron spectra.}

\FullConference{The XXVI International Symposium on Lattice Field Theory \\
		 July 14 - 19, 2008\\
		 Williamsburg, Virginia, USA}

\begin{document}

\section{Introduction}
 The main goal of the Hadron Spectrum Collaboration 
 is the calculation
 of low-lying excited hadron spectra. Exploratory calculations of both the 
 Nucleon \cite{AdamsThesis} and Delta \cite{theDeltas} spectra 
 have been completed
 as well as preliminary results on two-flavor lattices \cite{EricLatt08}. 
 These spectra may contain resonances as well as 
multi-particle states, both of which are distorted near decay thresholds. 
All-to-all propagators must also be used to create operators 
that interpolate multi-particle states, which consist of  
single particle states with finite momentum.
To extract a large number of 
both resonant and multi-particle states, a large variational basis of 
spatially extended operators must be used. The 
stochastic construction for all-to-all propagators results in significant 
computational savings  due to
a complete factorization of the source and sink information.
 The 
variance in stochastic estimates of all-to-all propagators may be reduced significantly by the dilution method 
\cite{TrinDil}, which is employed here. While this method has been 
demonstrated for mesons \cite{TrinDil} and simple multiparticle states \cite{Jimmy}, its effectiveness has not been established 
for the spatially extended baryon 
operators required by this project. This work is part of an ongoing effort \cite{Jimmy07} by the 
Hadron Spectrum Collaboration
to assess the utility of diluted all-to-all propagators in the extraction of 
low-lying baryon resonances. 

To this end, the diagonal correlators of three representative baryon
operators are calculated using various dilution schemes with a single set 
of stochastic sources. The relative error
on these correlators is then plotted as a function of the required 
number of Dirac 
matrix inversions. 
By examining the expected error falloff from including
additional noise sources, it is 
concluded that an increased level of dilution is preferable to 
multiple time-diluted noise sources.  However, a combination of 
dilution and an increase in stochastic noise sources may be the most efficient 
scheme.  
It is also observed that the fractional error for 
time + spin + color dilution is comparable to the point-to-all method. Furthermore, the 
fractional error on the \emph{exact} all-to-all is 
comparable to that of 
time + spin + color + spatial even-odd dilution.

\section{Methods}

\subsection{Stochastic Estimation}
The stochastic estimation of the quark propagator, $M^{-1}_{(\alpha a| \beta b)} (\textbf{x}, t|\textbf{x}_0, t_0)$
proceeds as follows: First, $N_r$ random sources, $\{\eta^{(r)}_{\alpha a} (\textbf{x}, t)\}$,  
are generated according to some
probability distribution.  The 
$Z_4 = \{1, -1, i ,-i\}$ distribution was used in this work, 
although similar results were obtained using both the $Z_2 = \{1, -1\}$
and $U(1) = \{e^{i\theta} \; | \; \theta \in (0, 2\pi] \}$ distributions. 
In the distributions described above, all elements are given equal 
probability. After the $N_r$ random sources have been generated, 
the linear 
system 
\begin{align}
		\nonumber
		M_{(\alpha a| \beta b)} (\textbf{x}, t|\textbf{x}', t')
		\; \phi^{(r)}_{\beta b} (\textbf{x}', t') =
    \eta^{(r)}_{\alpha a} (\textbf{x}, t)
		\end{align}
is solved for the $N_r$ solution vectors, 
$\{\phi^{(r)}_{\alpha a} (\textbf{x}, t)\}$. The quark propagator, 
$M^{-1}_{(\alpha a| \beta b)} (\textbf{x}, t|\textbf{x}_0, t_0)$, is 
given in terms of these source and solution vectors:  
 \begin{align}
		\nonumber
    M^{-1}_{(\alpha a| \beta b)} (\textbf{x}, t|\textbf{x}_0, t_0)
		 &= E \big[ \phi_{\alpha a} (\textbf{x}, t)
		 \; \eta^{*}_{\beta b} (\textbf{x}_0, t_0) \big]
		 \\ &\approx \frac{1}{N_r} \sum_r 
		 \phi^{(r)}_{\alpha a} (\textbf{x}, t) \; \eta^{*(r)}_{\beta b} (\textbf{x}_0, t_0).
\end{align}
The above quantity is an unbiased estimator of the quark propagator. 
However, in order to form unbiased estimators for hadronic observables, 
independent random noise sources must be used for each quark. For baryons in 
particular, $3N_r$ independent random sources must be created and used to form 
$3N_r$ solution vectors.

\subsection{Dilution}
The dilution method amounts to partitioning each of the random sources generated
above. For computational simplicity, a single set of noise sources was used in this work, although the dilution method can be used with any number of noise sources. 
Therefore rather than the $N_r$ random noise sources described above, 
a \emph{single} noise source, 
$\eta_{\beta b} (\textbf{x}_0, t_0)$, is partitioned according
to a complete set of $N_d$ orthogonal projectors, 
$\{P^{[d]}_{(\alpha a| \beta b)} (\textbf{x}, t|\textbf{x}', t')\}$. These 
projectors are applied
to form diluted noise sources
	\begin{align}
	\nonumber
		\eta_{\alpha a} (\textbf{x}, t) &= 
		\sum_{d = 1}^{N_d} \; 
		P^{[d]}_{(\alpha a| \beta b)} (\textbf{x}, t|\textbf{x}', t')
		\; \eta_{\beta b} (\textbf{x}', t') 
		\\\nonumber
		&= \sum_{d = 1}^{N_d} \;
		\eta^{[d]}_{\alpha a} (\textbf{x}, t).
	\end{align}
The $N_d$ diluted solutions are analogously obtained by solution of the 
following equation:
\begin{align}
		\nonumber
		M_{(\alpha a| \beta b)} (\textbf{x}, t|\textbf{x}', t')
		\; \phi^{[d]}_{\beta b} (\textbf{x}', t') =
    \eta^{[d]}_{\alpha a} (\textbf{x}, t).
		\end{align}

While any set of projectors satisfying the above properties may be used, 
simple `mask'-type projectors were employed for this work which correspond to
various coverings of the lattice. For example the `time' dilution scheme
corresponds to a set of projectors which each have support on a single 
timeslice only. This scheme can be partitioned further to form `time + spin' 
or `time + color' schemes, in which each projector has support on a single 
timeslice and spin or color, respectively. Spatial coverings of the lattice
may also be used. Projectors in the `time + spatial even-odd' scheme 
have support on a 
single timeslice and either the `odd' or `even' sites of the lattice, according
to a red-black checkerboarding scheme.

The dilution method may also be used in combination with exact 
solution of low-lying modes of the quark propagator \cite{TrinDil}. In this
case the diluted solutions are projected into the complement 
of the space spanned by the exactly solved low-lying modes. The solution
of low-lying modes is not used in this work. Furthermore, the use of 
a single time dilution projector reduces the required 
number of diluted sources 
to those with support on a single timeslice only. This also reduces 
the required number of Dirac matrix inversions and amounts to forming correlators 
without averaging over the source time. 
Additional time sources may be added for 
increased statistics.

\subsection{Implementation for Baryons}
The above method can be used to estimate baryonic two-point functions. 
This is done by forming the `source' and `sink' functions, 
$\Gamma^{[d_A d_B d_C]}_{\ell}(t)$ and 
$\Omega^{[d_A d_B d_C]}_{\ell}(t)$, each composed of
three $\eta$ or $\phi$ fields. These functions are given by 
\begin{align}\label{opEq}
\Gamma^{[d_A d_B d_C]}_{\ell}(t) = c^{(\ell)}_{\alpha \beta \gamma ; 
i j k} \sum_{\textbf{x}} \epsilon_{abc} \;
&\widetilde{\phi}^{(A)[d_A]}_{\alpha a i} (\textbf{x}, t)
\;
\widetilde{\phi}^{(B)[d_B]}_{\beta b j} (\textbf{x}, t)
\; \times
\\
\nonumber 
&\widetilde{\phi}^{(C)[d_C]}_{\gamma c k} (\textbf{x}, t)
\\\nonumber
\\\nonumber 
\Omega^{[d_A d_B d_C]}_{\ell}(t) = c^{(\ell)}_{\alpha \beta \gamma ; 
i j k} \sum_{\textbf{x}} \epsilon_{abc} \;
&\widetilde{\eta}^{(A)[d_A]}_{\alpha a i} (\textbf{x}, t)
\;
\widetilde{\eta}^{(B)[d_B]}_{\beta b j} (\textbf{x}, t)
\; \times
\\
\nonumber 
&\widetilde{\eta}^{(C)[d_C]}_{\gamma c k} (\textbf{x}, t)
\end{align}
where the $ c^{(\ell)}_{\alpha \beta \gamma ; 
i j k}$ coefficients represent both spin and covariant 
displacement structure for the $\ell$th operator and the $A,B,C$ indices 
represent three independent noise sources.
Although the operators described above have no spatial momentum,
this method can be trivially extended to operators with 
finite momentum by the addition of Fourier weights in the 
spatial sum. These functions are then 
combined via summation over the dilution indices ($[d_A d_B d_C]$) to form 
baryonic correlation functions.  

\begin{figure}
\centerline{
\raisebox{3mm}{\setlength{\unitlength}{1mm}
\thicklines
\begin{picture}(22,10)
\small
\put(8,6.5){\circle{6}}
\put(7,6){\circle*{2}}
\put(9,6){\circle*{2}}
\put(8,8){\circle*{2}}
\put(4,0){single-}
\put(5,-3){site}
\end{picture}}
\raisebox{3mm}{\setlength{\unitlength}{1mm}
\thicklines
\begin{picture}(22,10)
\small
\put(7,6.2){\circle{5}}
\put(7,5){\circle*{2}}
\put(7,7.3){\circle*{2}}
\put(14,6){\circle*{2}}
\put(9.5,6){\line(1,0){4}}
\put(4,0){singly-}
\put(2,-3){displaced}
\end{picture}}
\raisebox{3mm}{\setlength{\unitlength}{1mm}
\thicklines
\begin{picture}(22,12)
\small
\put(10,10){\circle{2}}
\put(4,10){\circle*{2}}
\put(16,10){\circle*{2}}
\put(10,4){\circle*{2}}
\put(4,10){\line(1,0){5}}
\put(16,10){\line(-1,0){5}}
\put(10,4){\line(0,1){5}}
\put(4,0){triply-}
\put(1,-3){displaced}
\end{picture}} }
\caption{The three types of elemental 
extended baryon operators used for this work. Solid circles 
represent smeared quark fields, lines represent smeared 
link fields, and hollow circles the location of the reference site.} 
\label{opsFig}
\end{figure}
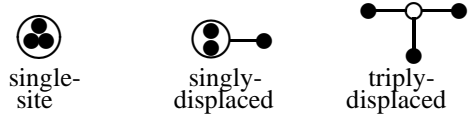

In order to access both radial and orbital excitations of baryons, 
operators which have an extended spatial structure must be used. This is 
achieved by the use of covariantly displaced operators in several geometries. These elemental
building blocks must be combined to form operators which transform according
to the lattice symmetries \cite{opConst}. To assess the effectiveness of 
various dilution schemes, a representative set of three operators was chosen 
consisting of a single-site, singly-displaced, and triply-displaced elemental
operator (Fig. \ref{opsFig}).
Contamination from higher-momentum modes is significantly reduced by using
operators which contain smeared quark fields. For this work,
a Gaussian smearing scheme was employed \cite{AdamsThesis} with a smearing
radius of $\sigma = 3.0$. The exponential smearing weights were approximated using 
$n_{\sigma} = 32$ iterations. Stout-link smearing \cite{stoutLink} has been 
shown \cite{AdamsThesis} to reduce noise in spatially extended operators. It 
was employed here with $n_{\rho}\rho = 2.5$ and $n_{\rho} = 16$. 

\section{Results}
\begin{figure}
\begin{center}
\includegraphics[width=.75\textwidth]{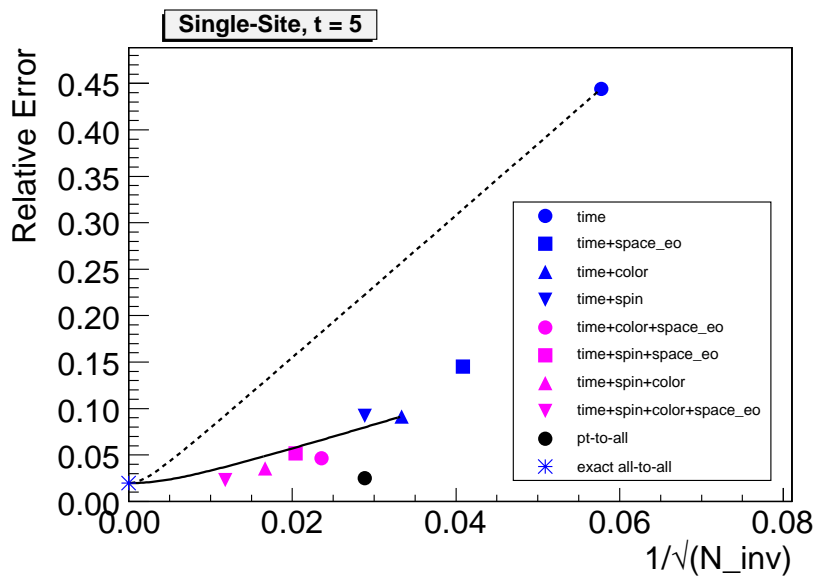}
\caption{The relative error of the diagonal correlator for a single-site 
elemental operator
(at $t=5$) 
versus the total number of Dirac matrix inversions required
for various dilution schemes with a single stochastic source. Along with these dilution schemes, the 
point-to-all result and the exact all-to-all result are plotted.
The dotted line represents the error falloff expected from increasing the
number of time-diluted noise sources. The solid line represents the
same for time+color-diluted noise sources.}
\label{SSfig}
\end{center}
\end{figure}

\begin{figure}
\begin{center}
\includegraphics[width=.75\textwidth]{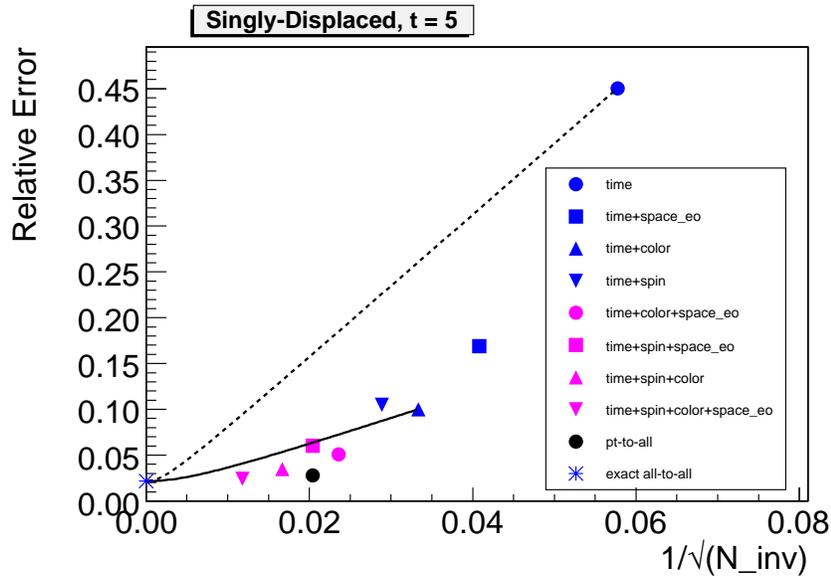}
\caption{The same as Fig. 2 for a singly-displaced elemental operator.}
\label{SDfig}
\end{center}
\end{figure}

\begin{figure}
\begin{center}
\includegraphics[width=.75\textwidth]{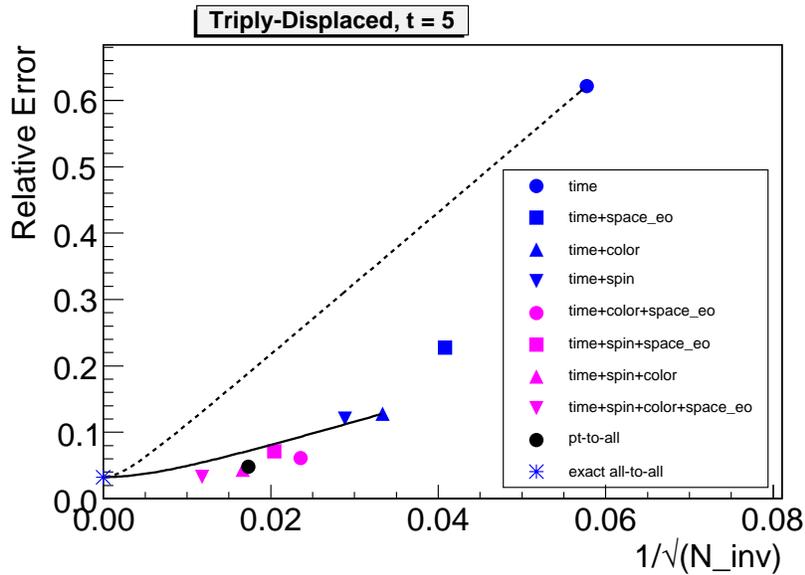}
\caption{The same as Fig. 2 for a triply-displaced elemental 
operator. The black point is slightly offset from the pink triangle for clarity.}
\label{TDTfig}
\end{center}
\end{figure}

To test the above formalism for the baryonic operators described above, 100 quenched gauge configurations were 
used with the following parameters: $L_s = 12$, $L_t = 48$, $a_s \approx .1 \; 
\mathrm{fm}$,
$\beta = 6.1$, and 
$m_{\pi} \approx 700 \; \mathrm{MeV}$.
The relative error of the diagonal correlator at $t=5$ was used as a basis for  
comparison while tests performed using the diagonal correlator at $t=10,15$ showed similar results. This quantity is plotted versus the total number of Dirac matrix 
inversions required (See Figs. \ref{SSfig}, \ref{SDfig}, and \ref{TDTfig}).  
The horizontal axis in these figures represents $1 / \sqrt{N_{inv}}$, where
$N_{inv}$ is the total number of required Dirac matrix inversions. In these 
coordinates, the error falloff
due to 
an increase in the 
number of noise sources for a fixed dilution scheme will be linear with a 
flattened approach to the gauge noise limit.
 
As all the subsequent points in these figures lie well below the dotted line, it is
clear that increasing the level of dilution past time dilution is more 
efficient than adding time-diluted noise sources. However, the 
improvement is less apparent for the solid line. Here there may be a small gain
in the fractional error, but computational cost and storage must be considered.
An increase in the number of dilution projectors by a factor $X$ results in an
$X^3$ increase in the number of components in the source and sink functions of 
Eq. \ref{opEq}. An increase in the number of diluted noise sources however, 
results in the same increase to the number of components of the source and sink
functions.  

\section{Conclusions}
The dilution method for variance reduction has been tested for baryons. As is 
seen in Figs. \ref{SSfig}, \ref{SDfig}, and \ref{TDTfig}, increasing the 
number of dilution projectors usually results in a decrease in error that is 
greater than the naive $1 / \sqrt{N}$ expected from increasing 
the number of diluted noise sources. It is also apparent from these figures
that time + spin + color dilution possesses comparable error to the 
point-to-all method and that time + spin + color + spatial even-odd dilution is
comparable to the exact all-to-all result.

While it is clear that increasing the level of dilution past time dilution 
is more efficient than adding more time-diluted sources, it 
may not be most efficient to employ dilution schemes finer than time + color dilution but rather increase the number of time + color-diluted noise sources. To complete these tests, simple fits to the diagonal correlators must be compared, as well as spatially extended meson operators. A new method \cite{newMeth} is
also being explored which will yield \emph{exact} all-to-all propagators for
a comparable cost. This method is based on a spectral decomposition of the 
quark smearing operator and will retain the desirable property of the 
factorization of the source and sink functions. 

\section{Acknowledgments}
This work was done with code written using the Chroma software suite \cite{chroma}. The authors would like to thank
Mike Peardon for valuable discussions about dilution methods. This work was supported by the U.S. National Science Foundation under the Award PHY-0653315 and
in part by Jefferson Science Associates, LLC under U.S. Dept. of Energy 
Contract No. DE-AC05-06OR23177.

\end{document}